\documentclass[aps,pra,showpacs,twocolumn,eqsecnum]{revtex4}
\usepackage{graphicx}
\usepackage{bm}
\begin{document}
\preprint{}

\title{Spin state in the propagation of quantum relativistic particles along classical trajectories}

\author{Jes\'{u}s Rubio and Alfredo Luis}
\email{alluis@fis.ucm.es}
\homepage{http://www.ucm.es/info/gioq}
\affiliation{Departamento de \'{O}ptica, Facultad de Ciencias
F\'{\i}sicas, Universidad Complutense, 28040 Madrid, Spain }

\date{\today}

\begin{abstract}
We address the propagation of the spin along classical trajectories for a 1/2-spin particle 
obeying the Dirac equation with scalar potentials. Focusing on classical trajectories as the 
exact propagation of wave-function discontinuities we find an explicit spin-transport law for 
the case of the Dirac oscillator. In the general case we examine the spin propagation 
along classical trajectories emerging as an approximation of the quantum dynamics via the 
mechanical analog of the optical eikonal asymptotic approach. Throughout we establish as 
many  parallels as possible with the equivalent situation for the electromagnetic field.
\end{abstract}


\maketitle

\section{Introduction}

In the theory of light is well known the deep relationship between electromagnetic and geometrical optics. 
Usually, geometrical optics is considered as an approximation valid in the limit of short wavelengths 
\cite{BW80,KO}. Furthermore, another radically different point of view is also possible: the light rays 
of geometrical optics determine the exact way in which the surfaces of field discontinuity propagate, for 
all wavelengths \cite{LU66,opt}. In this context, a natural question appears: what happens with the 
polarization? A simple calculus shows a coupling between the rays of the light, the gradient of the refraction 
index, and the field itself; consequently, it can be proved that the ellipse of polarization just rotates along 
the propagation, always keeping its shape \cite{LU66,opt}. It is a remarkable phenomenon that the polarization 
transformation has a purely geometrical origin since it depends only on the shape of the trajectory, becoming 
an example of topological dynamics.
 	
In this work we explore the behaviour of the spin state of a 1/2-spin particle when its spinorial
wave function is transported along classical trajectories, in the same sense we know 
how is the transport of the polarization for the photon. As in the optical case, we can follow two roadmaps.

On the one hand, we may regard the classical trajectories as the exact way in which any surface of discontinuity 
of the spinorial wave function must propagate \cite{AL}, as recalled in Sec. IIA. For 1/2-spin particles in purely 
scalar potentials, the conclusion of Ref. \cite{AL} was that the propagation of discontinuities provide no useful 
information about the spin transport along the classical trajectories. Thus we introduce an explicitly 
relativistic implementation of a classical scalar potential involving spin-dependent terms. This is the case of 
the Dirac oscillator \cite{MS}.  The corresponding propagation of spin discontinuities along the classical trajectories 
is carried out in Sec. IIB.
	
On the other hand, we may consider that classical trajectories also emerge in the quantum-mechanical 
version of the optical eikonal approach. This actually corresponds to the Wentzel-Kramers-Brillouin (WKB) 
methods \cite{WKB}, where the classical trajectories arise as an approximation of quantum dynamics 
valid for short values of the Planck constant. The spin propagation along these classical trajectories is carried 
out in Sec. III. To this end we mimic the optical case as far as possible. In order to exploit those parallels we 
recall the eikonal approach for Maxwell's equations in Appendix A.

\section{Quantum-discontinuity approach}

\subsection{Formalism}

For the sake of completeness, we recall here the basic tools required 
to address the propagation of discontinuities in optics as well as in
quantum mechanics, as presented in Ref. \cite{AL}. For both situations 
we will consider that the evolution is given by the solution of a system of linear 
partial differential equations 
\begin{equation}
\label{df}
\sum_{\mu,\nu} \frac{\partial }{\partial x^\mu} 
\left ( M^{\mu,\nu}_j \psi_\nu \right ) = 0, 
\end{equation}
where $x^\mu=x,y,z,t$ are the space-time coordinates,
$M^{\mu,\nu}_j$ are functions of $x^\eta$, and 
$\psi_\nu (x^\mu)$ are the Cartesian components 
either of a spinorial quantum wave function or of 
a classical electromagnetic field. 

As discussed in Ref. \cite{LU66}, the 
Eqs. (\ref{df}) are conditions for the components 
$\psi_\nu$ in every point where they are 
continuous, but they cannot establish conditions  
for the boundary values of $\psi_\nu$ on a surface 
of discontinuity. Therefore, in order to deal with
discontinuities it is advantageous to replace 
Eqs. (\ref{df}) by their integral counterparts. To this 
end we consider volume integrals of Eq. (\ref{df}), 
that can be then suitably converted into surface 
integrals using the divergence theorem \cite{LU66}
\begin{equation}
\label{sf}
\int_\Gamma d \Gamma \sum_{\mu,\nu} 
\frac{\partial }{\partial x^\mu} 
\left ( M^{\mu,\nu}_j \psi_\nu \right ) =
\sum_{\mu,\nu} \int_\gamma d \gamma_\mu 
M^{\mu,\nu}_j \psi_\nu =0 , 
\end{equation}
where $d \Gamma = dxdydzdt$ is the differential of four-dimensional volume and $d \gamma_\mu$ 
are the Cartesian components of the surface 
element normal to the three-dimensional 
surface $\gamma$ enclosing $\Gamma$. The last 
equality in Eq. (\ref{sf}) is fully equivalent 
to Eq. (\ref{df}) when $\psi_\nu$ are continuous. 
On the other hand Eq. (\ref{sf}) is more general 
since it can be applied without difficulties 
when $\psi_\nu$ are discontinuous.

Our aim is to derive conditions for the discontinuities of $\psi_\nu$ by imposing 
Eq. (\ref{sf}). Denoting by $S (\bm{x},t)=0$ the surface of discontinuity, we apply Eq. (\ref{sf}) 
to two volumes $\Gamma_1$, $\Gamma_2$ connected by $S$, as well as to the whole 
volume $\Gamma_1 +  \Gamma_2$ (see Fig. 1). This leads to 
\begin{equation}
\label{ff}
\sum_{\mu,\nu} [ \psi_\nu ] \frac{\partial S}
{\partial x^\mu} M^{\mu,\nu}_j = 0,
\end{equation}
where for each  $\bm{x},t$ we have that $[ \psi_\nu ]$ denotes the difference 
between the boundary values of $\psi_\nu$ at the two sides of $S$, i. e.,
$[ \psi_\nu]  =  \psi_\nu ( \mathrm{at} \, \Gamma_2 ) - \psi_\nu ( \mathrm{at} \, \Gamma_1) $. 
We have considered $d \gamma_\mu \propto \partial S / \partial x^\mu$ and that 
$\Gamma_{1,2}$ are arbitrary. Moreover, we assume that $M^{\mu,\nu}_j$ are 
continuous at $S$. Equations (\ref{ff}) are the conditions we were looking for. 

\begin{figure}
\includegraphics[width=5cm]{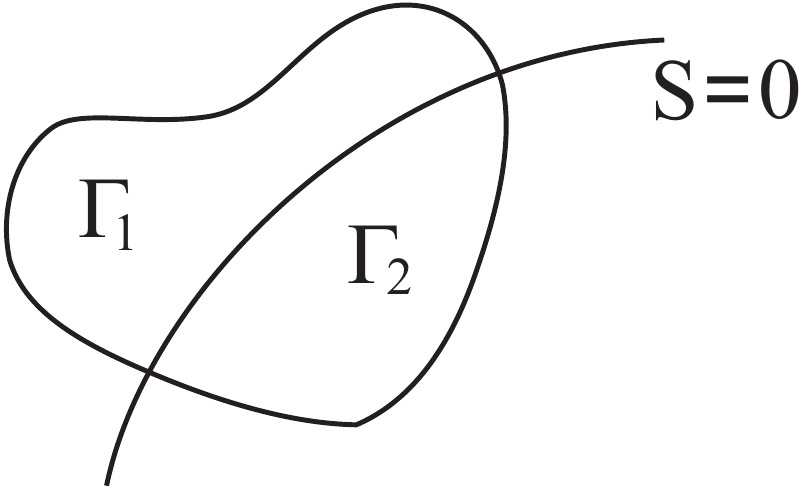}
\caption{Diagram illustrating the volumes 
$\Gamma_1$ and $\Gamma_2$ joined by the 
surface of discontinuity $S=0$.}
\end{figure}

In particular, when this formalism is applied to the Dirac equation with scalar potential $V(\bm{x})$ 
and definite fixed energy $E$, i. e., for harmonic wave functions, we get \cite{AL}
\begin{equation}
\label{rl}
E = \sqrt{c^2 \left ( \bm{\nabla} W \right )^2 + m^2 c^4}+ V  ,
\end{equation} 
having introduced a separation of variables such that $S (\bm{x},t)=W(\bm{x})-Et$. This means that 
the surface $S$ where the discontinuity of harmonic wave functions takes place must satisfy the 
classical Hamilton-Jacobi equation for the relativistic Hamiltonian $H = \sqrt{c^2 \bm{p}^2 + m^2 c^4} 
+  V$. Therefore, the evolution of $S$ is fully determined by the classical trajectories for the problem 
\cite{SC71}, so we say that $S$ follows the classical trajectories. The result is exact since no approximation 
nor limiting procedure whatsoever have been used.

\subsection{Dirac oscillator}

It is worth noting that for scalar potentials the formalism outlined above provides no information 
about the spin transport along the classical trajectories. As mentioned in the introduction,
we will address this issue within the quantum-discontinuity approach to classical trajectories 
considering the relativistic formulation of scalar potentials including spin-dependent terms. 
This is the case of the Dirac oscillator \cite{MS}.
 
The Dirac oscillator is a simple model for a relativistic isotropic oscillator of frequency $\omega$ and 
mass $m$,  that preserves the fully linear character of the Dirac equation 
 \begin{equation}
 \label{Do}
 i \hbar  \frac{\partial \bm{\Psi}}{\partial t} = c \left ( \bm{p} \cdot \bm{\alpha}- i m \omega \bm{x} \cdot \bm{\alpha} 
 \beta + mc \beta  \right ) \bm{\Psi} ,
 \end{equation}
 where $\bm{p} = - i \hbar  \bm{\nabla}$,  $\bm{\alpha}$ and $\beta$ are the $4 \times 4$ matrices
\begin{equation}
\label{si}
\bm{\alpha} = \pmatrix{0 & \bm{\sigma} \cr  \bm{\sigma} & 0} , \quad 
\beta = \pmatrix{I & 0\cr  0 & - I},
\end{equation}
$ \bm{\sigma}$ are the three Pauli matrices, $I$ the $2 \times 2$ identity, and the spinorial 
four-dimensional wave function $\bm{\Psi}$ depends on the Cartesian coordinates $\bm{x}$ 
and time $t$.

At this stage we cannot apply directly the previous formalism because the lack of partial derivatives 
at the terms $( - i m \omega \bm{x} \cdot \bm{\alpha} \beta + mc \beta ) \bm{\Psi}$ impedes to express Eq. 
(\ref{Do}) on the form (\ref{df}). As in classical optics we can avoid this difficulty by restricting our analysis 
to time-harmonic wave functions for which
\begin{equation}
\bm{\Psi} = \frac{i\hbar}{E} \frac{\partial \bm{\Psi}}{\partial t}  ,
\end{equation}
so that Eq. (\ref{Do}) can be rewritten as 
\begin{equation}
 \frac{\partial \bm{\Psi}}{\partial t} = -  c \bm{\alpha} \cdot  \bm{\nabla} \bm{\Psi} + \frac{c}{E}
 \frac{\partial}{\partial t}  \left [ \left (  - i m \omega \bm{x} \cdot \bm{\alpha} \beta + mc \beta  \right ) 
 \bm{\Psi} \right ] .
 \end{equation}
Since all terms have now partial derivatives we can apply Eq. (\ref{ff}) leading to 
\begin{equation}
 [ \bm{\Psi} ] \frac{\partial S}{\partial t} = -   c \bm{\nabla} S \cdot \bm{\alpha} [ \bm{\Psi}] + \frac{c}{E} 
 \frac{\partial S}{\partial t} \left (  - i m \omega \bm{x} \cdot \bm{\alpha} \beta + mc \beta  \right ) [ \bm{\Psi} ] ,
 \end{equation}
where $ [ \bm{\Psi} ] $ represents again the difference between the boundary values of $\bm{\Psi}$ at the 
two sides of a certain surface $S(\bm{x},t)$ where such discontinuity takes place.
 
For the sake of clarity, this four-dimensional equation can be split into a pair of two-dimensional equations using Eq. (\ref{si})
\begin{eqnarray}
 \label{Dod}
 & \left ( mc^2 - E \right )  \frac{\partial S}{\partial t} [ \bm{\phi} ] = c \left ( E \bm{\nabla} S 
 - i  \frac{\partial S}{\partial t} m \omega \bm{x} \right ) \cdot \bm{\sigma} [\bm{\chi} ] , & \nonumber \\
  & & \nonumber \\
  & -  \left ( mc^2 + E \right )  \frac{\partial S}{\partial t} [ \bm{\chi}  ] = c \left ( E \bm{\nabla} S 
 + i  \frac{\partial S}{\partial t} m \omega \bm{x} \right ) \cdot \bm{\sigma} [\bm{\phi} ]  , & \nonumber \\
  & & 
 \end{eqnarray}
where the two-dimensional spinors $\bm{\phi}$, $\bm{\chi}$ are defined as 
\begin{equation}
\label{ft}
\bm{\Psi} = \pmatrix{ \bm{\phi}  \cr   \bm{\chi}  } .
\end{equation}
Using one of the Eqs. (\ref{Dod}) to remove the discontinuity of the lower spinor $ [ \bm{\chi} ]$, and taking
into account  the general relation
\begin{equation}
\label{vr}
\left (\bm{A} \cdot \bm{\sigma} \right )\left (\bm{B} \cdot \bm{\sigma} \right ) = 
 \left (\bm{A} \cdot \bm{B} \right ) I + i \left ( \bm{A} \times \bm{B} \right ) \bm{\sigma},
 \end{equation}
 
\begin{widetext}

we get a two-dimensional equation for the discontinuity of the upper spinor  $ [ \bm{\phi} ]$

\begin{equation}
\label{Doee}
\left (   \frac{\partial S}{\partial t}  \right )^2  \left (E^2 -  m^2 c^4  \right )[ \bm{\phi}] = 
c^2 \left [ E^2 \left ( \bm{\nabla} S \right )^2 + \left (   \frac{\partial S}{\partial t}  \right )^2 m^2 \omega^2 \bm{x}^2 
+ 2 m E \omega  \frac{\partial S}{\partial t}  \left ( \bm{x} \times \bm{\nabla} S \right ) \cdot \bm{\sigma} \right ]
[\bm{\phi}] .
\end{equation}

This equation is of the form 
\begin{equation}
\label{sDo}
\left ( \bm{x} \times \bm{\nabla} S \right ) \cdot \bm{\sigma} [\bm{\phi}] = \lambda [\bm{\phi}] ,
\end{equation}
so that $\lambda = \pm |  \bm{x} \times \bm{\nabla} S  |$ . Hence we must have

\begin{equation}
\label{HJDo}
\left (   \frac{\partial S}{\partial t}  \right )^2  \frac{E^2 -  m^2 c^4}{E^2 c^2} - \left ( \bm{\nabla} S \right )^2 
- \frac{1}{E^2} \left (   \frac{\partial S}{\partial t}  \right )^2 m^2 \omega^2 \bm{x}^2 \pm  \frac{2 m \omega}{E} 
\frac{\partial S}{\partial t} \left | \bm{x} \times \bm{\nabla} S \right | = 0 .
\end{equation}

\end{widetext}

This is the equation that $S( \bm{x},t)$ must satisfy in order to describe a surface of discontinuity for the 
quantum wave function of a Dirac oscillator. Note that actually there is no sign freedom in Eq. (\ref{HJDo}) 
because there is no  sign ambiguity in Eq. (\ref{Doee}).  

Since the Hamiltonian we are considering is time-independent and the wave function is harmonic, 
it is natural to use the method of separation of variables to separate out the time in the form $S(\bm{x},t) = W(\bm{x})-Et$, 
leading to a much simpler equation for $W(\bm{x})$ 
\begin{equation}
\label{HJDo2}
 \frac{E^2 -  m^2 c^4}{c^2} - \left ( \bm{\nabla} W \right )^2  - m^2 \omega^2 \bm{x}^2 \mp  2 m \omega \left | \bm{x} 
 \times \bm{\nabla} W  \right | = 0 .
\end{equation}

Equations (\ref{HJDo}) and (\ref{HJDo2}) can be readily interpreted as the Hamilton-Jacobi for the Dirac 
oscillator, and its solutions are the corresponding classical trajectories where $\bm{\nabla} W$ is normal 
to the discontinuity surface and tangent to the trajectory at each point. These trajectories are the same 
plane ellipses of the standard oscillator since the extra term $\bm{x} \times \bm{\nabla} W$ just depends 
on the orbital angular momentum that is a constant of the motion.

The key point for our purposes here is that the eigenvalue equation (\ref{Doee}) determines a single 
definite spin state at each point so it explicitly contains how the spin is transported along the classical trajectories. 
More specifically, the spin state is always an eigenstate of the spin projection along the direction of the orbital 
angular momentum $ \bm{x} \times \bm{\nabla} W $, i. e., normal to the plane where the trajectory is contained
(see Fig. 2). Thus we may say that the spin does not influence on the trajectory while the trajectory forces the 
spin state. Therefore, it seems that the topological properties of the photon-polarization transport are lost in the 
1/2-spin case.

\begin{figure}
\includegraphics[width=4cm]{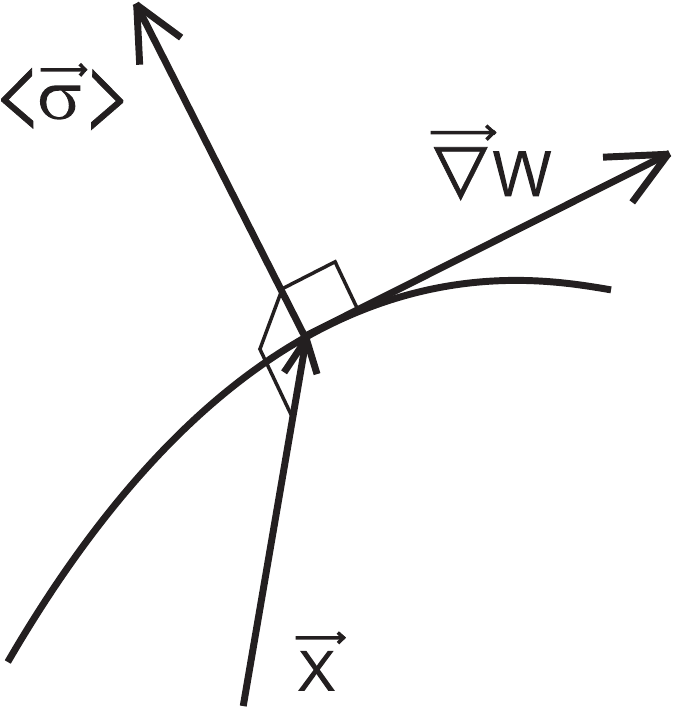}
\caption{Diagram illustrating the relation between the tangent  $\bm{\nabla} W$ to the trajectory $\bm{x}$,  
and the spin state represented by the vector $\langle \bm{\sigma} \rangle$, which is always orthogonal 
both to  $\bm{\nabla} W$ and $\bm{x}$.}
\end{figure}

Finally we stress that this explicit spin evolution for the discontinuity holds because of the presence 
of a term proportional to $ ( \bm{x} \times \bm{\nabla} W ) \cdot \bm{\sigma}$ in (\ref{Doee}), which is a strong 
spin-orbit coupling that is also preserved in the the non-relativistic limit \cite{MS}. Nevertheless,  the physical 
system being described is an harmonic oscillator whose potential is scalar.
 
 \section{Eikonal approach}
 
The Dirac oscillator allowed us to determine the evolution of the spin state by applying the
formalism of the propagation of discontinuities with no approximation. However, for physical situations 
with purely scalar potentials the divergence theorem does not provide enough information to determine 
the spin evolution along trajectories. To avoid this  obstacle we consider the emergence of classical 
trajectories as the quantum-mechanical version  of the optical eikonal. This is the WKB method 
\cite{WKB}. In this work we are particularly interested on applying this method to determine the 
transport of the spin along  such classical trajectories. We put particular emphasis in preserving as far as 
possible the parallelism with the electromagnetic case, in particular regarding its simplicity.
 
 \subsection{Transport equation}
  
Let us start with the Dirac equation for a 1/2-spin particle within an arbitrary scalar potential $V (\bm{x} )$, 
with definite energy $E$, mass $m$ and a time-harmonic wave function 
 \begin{equation}
 \label{D}
 \left [ - i \hbar c \bm{\alpha} \cdot \bm{\nabla} + mc^2 \beta + V (\bm{x} ) \right ] \bm{\Psi}  =E  \bm{\Psi} .
 \end{equation}
 
 Instead of considering spin-dependent potentials, now we address a more transparent approach 
 just in terms of $V (\bm{x} )$, mimicking as far as possible the electromagnetic 
 situation where light evolution is governed by scalar position-dependent quantities $\epsilon (\bm{x}), 
 \mu (\bm{x})$.
  
Splitting the four-dimensional spinor $\bm{\Psi}$ as in Eq. (\ref{ft}), and using Eq. (\ref{si}), the 
four-dimensional matrix equation (\ref{D})  can be decoupled into a pair of  two-dimensional equations, 
as follows
\begin{eqnarray}
 \label{Dd}
 & - i \hbar c \left (  \bm{\sigma} \cdot \bm{\nabla} \right ) \bm{\chi}  = \left ( E - m c^2 - V \right ) \bm{\phi} , & 
 \nonumber \\
  &- i \hbar c \left (  \bm{\sigma} \cdot \bm{\nabla} \right )  \bm{\phi}  = \left ( E + m c^2 - V \right ) \bm{\chi}  , &
 \end{eqnarray}
These two equations share {grosso modo the structure of the second pair of Maxwell's equations in 
Eq. (\ref{Me2}), via the rough correspondences  
\begin{equation}
\hbar \leftrightarrow 1/k_0 , \quad \bm{\phi}, \bm{\chi} \leftrightarrow \bm{E}, \bm{H}, \quad 
V(\bm{x}) \leftrightarrow \mu (\bm{x}), \epsilon (\bm{x}) .
\end{equation}
However, in the Dirac's case there is no counterpart of the first two Maxwell equations (\ref{Me1}). 
This is to say that matter waves lack the transversality condition satisfied by light waves, so we may 
expect some basic differences regarding the vectorial behavior.

 Continuing with the analogy, we construct now a kind of second-order differential wave equation for the 
 Dirac case equivalent to Eq. (\ref{we}). This  is always possible clearing $\bm{\chi}$ from the second 
 equation of  Eq. (\ref{Dd})  and substituting it into the first one 
 
 \begin{widetext}
 
 \begin{equation}
 \label{phi}
 \hbar^2 \bm{\nabla}^2 \bm{\phi} + \left [ \left ( \frac{E- V}{c} \right )^2-m^2 c^4 \right ] \bm{\phi} + 
 \frac{\hbar^2 \bm{\nabla} V  \cdot \bm{\nabla} }{E- V+mc^2} \bm{\phi}  + i  \frac{\hbar^2  \bm{\sigma} 
 \cdot \left [ \bm{\nabla} V  \times \bm{\nabla} \right ]}{E- V+mc^2} \bm{\phi} = 0.
 \end{equation}
 
\end{widetext}

Going further into this development, we apply  the eikonal approximation $\hbar \rightarrow 0$ as an 
analog of the electromagnetic Eq. (\ref{expanE}), looking for  solutions of the form 
\begin{equation}
\label{expan}
\bm{\phi} = \left ( \bm{\phi}_0 + \hbar \bm{\phi}_1 + \ldots \right ) e^{i \tilde{S} / \hbar} ,
\end{equation}
where  $\tilde{S} (\bm{x} ,t)$ is the phase of the wave function. We are using the same letter 
used above for a surface of discontinuity for the reasons shown below. Substituting Eq. (\ref{expan}) 
into Eq. (\ref{phi}) and considering again the separation of variables $\tilde{S} (\bm{x} ,t) = \tilde{W} 
(\bm{x} )-Et$ for a time-independent Hamiltonian and harmonic wave functions,
we get the following result for the order $\hbar^0$, after some lengthy but straightforward algebra
\begin{equation}
\label{Wf}
\left [ \nabla \tilde{W} (\bm{x} ) \right ]^2 =\left [ \frac{E- V(\bm{x})}{c} \right ]^2-m^2 c^4 .
\end{equation}
This is the analog of an eikonal equation independent of spin for the electron, fully analogous to the 
Eq. (\ref{eik}) of the electromagnetic case. Comparing Eqs. (\ref{Wf}) and (\ref{eik}), it is very remarkable 
that the only real difference between them is the fact of the former has mass. 

Note that this eikonal approach lead us to an equation for the phase $\tilde{S}$ that is formally identical 
to  the classical Hamilton-Jacobi equation for the relativistic Hamiltonian (\ref{rl}). So we may say that 
$\tilde{S}$ represents the classical action of the system. Moreover, note that the phase $\tilde{S}$ here 
and the surface of discontinuity $S$ in Sec. II obey exactly the same dynamics as expressed by 
Eqs. (\ref{rl}) and (\ref{Wf}), so deep down they represent the same physical entity but emerging under 
slightly different approaches. This is the reason why we use closely related symbols for both quantities.  
Moreover, from now on we take $\tilde{W}= W$ for the sake of simplicity.

Roughly speaking, the equivalence between $\tilde{S}$ and $S$ holds since when $\hbar \rightarrow 0$ 
the phase $\tilde{S}/\hbar$ of the quantum wave function in Eq. (\ref{expan}) becomes effectively discontinuous 
at every point of the space by displaying  very large variations in very short displacements. 

On the other hand, we now introduce the expansion (\ref{expan}) in Eq. (\ref{phi}), and again after some 
lengthy but straightforward algebra, the order $\hbar^1$ of Eq. (\ref{phi}) is
\begin{equation}
 \label{gte}
\frac{d}{ds} \bm{\phi}_0 = - \frac{\bm{\nabla}^2 W}{2 | \bm{\nabla} W | } \bm{\phi}_0
- \frac{\left (\bm{\nabla} V \cdot \bm{\sigma} \right ) \left (\bm{\nabla} W \cdot \bm{\sigma} \right ) }
{2 \left ( E + mc^2-V \right )  | \bm{\nabla} W | } \bm{\phi}_0 ,
 \end{equation}
where we have used that the derivation along the arc length $s$ of the trajectory reads
\begin{equation}
\frac{d}{ds} = \frac{ \bm{\nabla} W \cdot \bm{\nabla}}{\left | \bm{\nabla} W \right | } .
\end{equation}
After relation (\ref{vr}), this is equivalent to

 \begin{widetext}

\begin{equation}
 \label{mgte}
\frac{d}{ds} \bm{\phi}_0 = - \frac{\bm{\nabla}^2 W}{2 | \bm{\nabla} W | } \bm{\phi}_0
- \frac{\bm{\nabla} V \cdot \bm{\nabla} W}{2 \left ( E + mc^2-V \right )  | \bm{\nabla}W | } 
\bm{\phi}_0 - i \frac{\left ( \bm{\nabla}V \times \bm{\nabla}W \right ) \cdot \bm{\sigma}}{2 
\left ( E + mc^2-V \right )  | \bm{\nabla} W | } \bm{\phi}_0 .
 \end{equation}
 
  \end{widetext}

This is precisely the transport equation we were looking for. As we announced, it shares exactly the 
same structure of the electromagnetic counterpart in  Eq. (\ref{et}) since both are linear equations of 
the form $d \bm{A} /ds = M \bm{A}$. How the physical objects are coupled in each case is, however, 
different. In the Maxwell transport, the electromagnetic field is a pair of three-dimensional vectors 
coupled with the gradient of the inhomogeneous parameters $\bm{\nabla} \epsilon, \bm{\nabla}\mu$, 
as well as with the tangent to the trajectory $\bm{\nabla} L$. These are all three-dimensional 
vectors. Meanwhile, Dirac transport couples vectorially the tangent to the trajectory $\bm{\nabla} W$},  
the gradient of the scalar potential $\bm{\nabla} V$, and the spin $\bm{\sigma}$. However, now the 
fermionic field, which is divided into two bispinors, has a spinorial four-dimensional character, so it 
cannot be specifically coupled on the same grounds with the other dynamical objects $\bm{\nabla} V$ 
and $\bm{\nabla} W$, which are three-dimensional vectors. 

Note that in the non relativistic limit $E-mc^2 \ll mc^2$ the spin-trajectory coupling between 
$\bm{\sigma}$ and $\bm{\nabla} W$ disappears. Thus spin transport in the absence of magnetic 
fields is a purely relativistic phenomenon.

\subsection{Amplitude and spin}
 
The general transport equation  (\ref{mgte}) can be split into  equations for amplitude $ | \bm{\phi}_0 (\bm{x}) |$ 
and local spin state $\bm{u}_0 (\bm{x})$ after decomposing $\bm{\phi}$ in the form 
\begin{equation}
\bm{\phi}_0 = | \bm{\phi}_0  | \bm{u}_0 , \qquad | \bm{u}_0  | = 1. 
\end{equation}
For the amplitude $| \bm{\phi}_0  |$ we get  
\begin{equation}
 \label{}
\frac{d}{ds} | \bm{\phi}_0  |  = - \left ( \frac{\bm{\nabla}^2 W}{2 | \bm{\nabla} W | }  + \frac{\bm{\nabla} V \cdot 
\bm{\nabla} W}{2 \left ( E + mc^2-V \right )  | \bm{\nabla} W | }
\right ) | \bm{\phi}_0  |  ,
\end{equation}
while for the spin state $ \bm{u}_0 $ we have
\begin{equation}
 \label{dsu0}
\frac{d}{ds} \bm{u}_0 = - i \frac{\left (\bm{\nabla} V \times \bm{\nabla} W \right ) \cdot \bm{\sigma}}{2
 \left ( E + mc^2-V \right )  | \bm{\nabla} W | } \bm{u}_0 .
 \end{equation}
 In particular this last expression allow us to derive a transport equation for the local mean value 
 $\langle A \rangle$ of any spin observable $A$ as
 \begin{equation}
 \langle A \rangle = \bm{u}_0^\dagger A \bm{u}_0 , \qquad
 \frac{d}{ds}   \langle A \rangle \propto -i \langle [ \left ( \bm{\nabla} V \times \bm{\nabla} W \right ) 
 \cdot \bm{\sigma}, A ] \rangle ,
 \end{equation}
 and in particular
 \begin{equation}
\frac{d}{ds} \langle \bm{\sigma} \rangle \propto \langle \bm{\sigma} \rangle \times 
\left ( \bm{\nabla} V \times \bm{\nabla} W \right ) .
\end{equation}
By local spin mean values we mean that  $ \langle A \rangle $ depends on $\bm{x}$. This is 
that the spin transport is made of consecutive local rotations around the vector $\bm{\nabla} V \times 
\bm{\nabla} W$.  In particular,  the projection of the spin on the vector  $\bm{\nabla} V \times 
\bm{\nabla} W$ is constant $\langle \left ( \bm{\nabla} V \times \bm{\nabla} W \right )  \cdot \bm{\sigma} 
\rangle = \mathrm{constant}$ (see Fig. 3).  

\begin{figure}
\includegraphics[width=6cm]{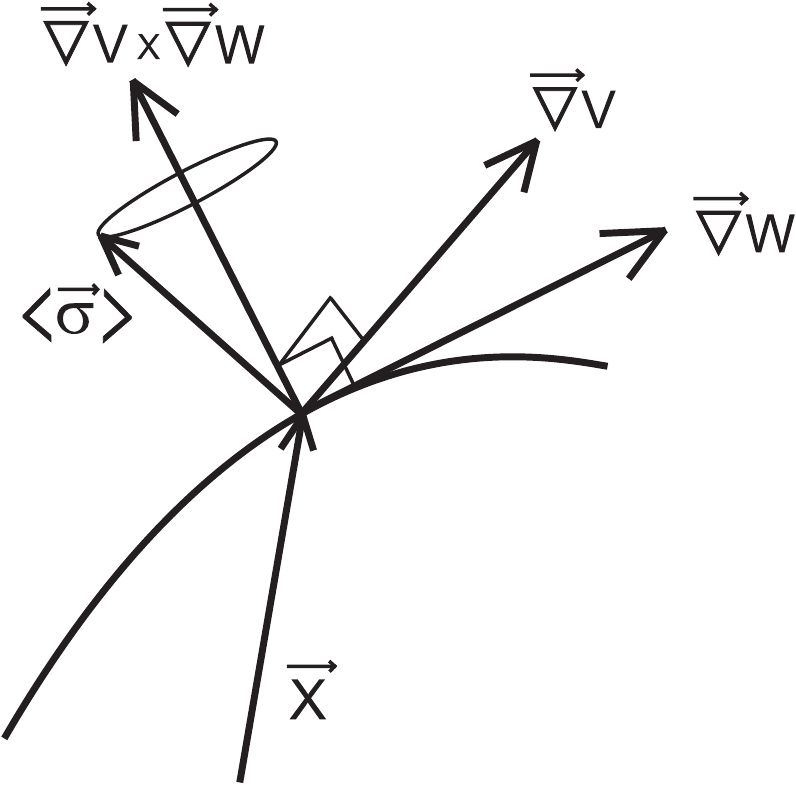}
\caption{Diagram illustrating the evolution of the spin state represented by the vector $\langle \bm{\sigma} 
\rangle$ rotating around the axis  $\bm{\nabla} V \times \bm{\nabla} W$ where $\bm{\nabla} W$ is tangent 
to the trajectory $\bm{x}$ and $\bm{\nabla} V$ is the gradient of the potential.}
\end{figure}

We can particularize to the harmonic oscillator $V(\bm{x} ) \propto \bm{x}^2$ in order to compare with the 
result for the Dirac oscillator above. In such a case the trajectories are in the plane defined by 
$\bm{\nabla} V$ and $\bm{\nabla} W$. Nevertheless, at difference with the Dirac-oscillator case
the spin state needs not be constant and orthogonal to the trajectory plane, and its evolution will 
depend in general of the trajectory followed. Moreover, the transport equation depends on dynamical 
features other than the form of the trajectory, so seemingly also in this case the geometrical character of 
the transport is lost in the transition from light to matter. 

\section{Applications}

We explore now the application of the above formalism. The most trivial example is the case of a 
free particle. In such a case $\bm{\nabla} V =\bm{0}$ so that the spin is constant along the trajectory. 
Thus, to obtain simple illustrative and nontrivial results we focus on two further examples: 
planar circular and circular helicoidal trajectories. In particular, the second one admits a direct comparison 
with similar applications of the optical eikonal regarding the emergence of geometric phases \cite{opt}.

\subsection{Planar circular trajectory}

The simplest nontrivial example is provided by a planar circular trajectory for a particle moving in a 
three-dimensional central potential $V = V ( \sqrt{x^2+y^2+z^2} )$, with 
\begin{equation}
x= r_0 \cos \left (  s / r_0 \right ) , \quad  y= r_0 \sin \left (s / r_0 \right ) , \quad z = 0,
\end{equation}
where $s$ is the arc-length parameter, being $r_0$ the radius of the trajectory. The corresponding 
tangent vector is
\begin{equation}
\frac{\bm{\nabla} W}{\left | \bm{\nabla} W \right |} =  \left ( -  y,  x, 0 \right ) /r_0 ,
\end{equation} 
while $\bm{\nabla} V = - k (x,y,0)$, for a suitable constant $k$ depending in general on $r_0$.
Therefore, the spin transport equation (\ref{dsu0}) becomes
\begin{equation}
\frac{d}{ds} \bm{u}_0 =  i  \frac{k r_0}{2 \left ( E + mc^2-V_0 \right ) } \sigma_z  \bm{u}_0 ,
 \end{equation}
 where $V_0 = V (r_0 )$ . The solution is rather simple 
\begin{equation}
\label{err}
\bm{u}_0 (s) = e^{ i \theta \sigma_z } \bm{u}_0 (0) , \quad \theta = \frac{k r_0 s}{2 \left ( E + 
mc^2-V_0 \right ) } .
\end{equation}
The result is a spin rotation around the axis $z$ of angle $\theta$ (see Fig. 4). 

\begin{figure}
\includegraphics[width=6cm]{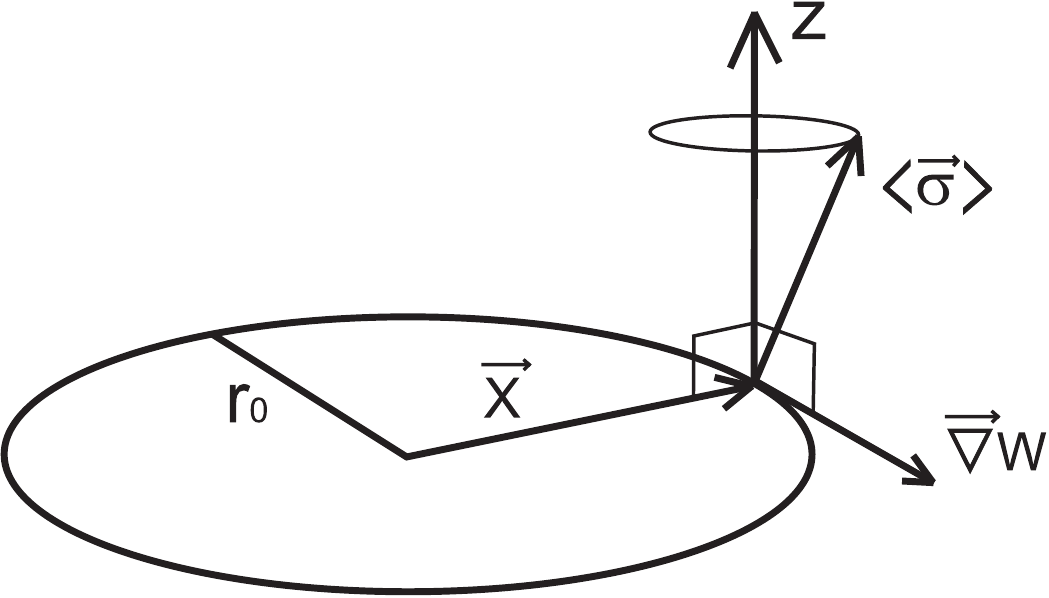}
\caption{Diagram illustrating the evolution of the spin state represented by the vector $\langle \bm{\sigma} 
\rangle$ rotating around the axis  $z$ normal to the plane of the trajectory.}
\end{figure}

It is worth noting that when the particle completes a revolution of arc-length $s_0 = 2 \pi r_0$  the 
spin does not return to its initial state $\bm{u}_0 (s_0) \neq \bm{u}_0 (0)$. This would be a kind of 
geometrical phase if it were not by the fact that  the phase acquired in closed loops does not depend 
only on the form of the trajectory, but also on dynamical factors represented by $k$ and $V_0$. Next 
we discuss this point with a more elaborated example.
 
\subsection{Circular helicoidal trajectory}

Let us consider a circular helicoidal trajectory  
\begin{equation}
x= r_0 \cos \left ( \Omega z \right ) , \quad y= r_0 \sin \left ( \Omega z \right ) , 
\end{equation}
where the arc-length is $s= \sqrt{1 + \Omega^2 r_0^2}$,  $r_0$ is the radius, and the helix pitch is 
$2 \pi/ \Omega$. The tangent  to the trajectory reads 
\begin{equation}
\frac{\bm{\nabla} W}{\left | \bm{\nabla} W \right |} = 
\frac{\left ( - \Omega  y,  \Omega x, 1 \right )}{ \sqrt{1 + \Omega^2 r_0^2}} .
\end{equation} 
We assume that this motion in the plane $x,y$ takes place in a two-dimensional central potential 
$V = V (\sqrt{x^2+y^2} )$, so that $\bm{\nabla} V = - k (x,y,0)$, for a suitable constant $k$ that 
depends on $r_0$. Otherwise the particle is free to move along axis $z$ which is the helix axis. 
Taking into account that
\begin{equation}
\frac{\bm{\nabla} V \times \bm{\nabla} W}{\left | \bm{\nabla} W \right |} = 
\frac{k \left ( -  y,  x, \Omega r_0^2 \right )}{ \sqrt{1 + \Omega^2 r_0^2}} ,
\end{equation} 
the spin transport equation (\ref{dsu0}) becomes
\begin{equation}
\frac{d}{ds} \bm{u}_0 = i \mu \left ( y \sigma_x - x \sigma_y + \Omega r_0^2 \sigma_z \right ) \bm{u}_0 ,
 \end{equation}
 where 
 \begin{equation}
 \mu = \frac{k/\sqrt{1+\Omega^2 r_0^2}}{2 \left ( E + mc^2-V_0 \right ) } , \qquad V_0 = V (r_0 ) .
 \end{equation}
This can be easily solved using standard techniques of this kind of problems, leading to 
\begin{equation}
\label{er}
\bm{u}_0 (s) = e^{- i \delta \sigma_z} e^{i \left [ \left ( \delta + \theta \right ) \sigma_z -
 \varphi \sigma_y \right ] } \bm{u}_0 (0)  ,
\end{equation}
with 
\begin{equation}
\delta = \frac{\Omega s}{2 \sqrt{1 + \Omega^2 r_0^2}}, \quad \theta = \mu r_0^2 \Omega s, 
\quad \varphi = \mu r_0 s /2 .
\end{equation}
For all practical purposes the rest-mass energy is the largest factor so that $E + mc^2 - V_0 
\simeq 2 mc^2$. Therefore
\begin{equation}
 \mu \simeq \frac{k/\sqrt{1+\Omega^2 r_0^2}}{2 mc^2 } ,
 \end{equation}
and we can regard $\mu$ as well as $\theta$ and $\varphi$ as very small parameters.
This allows a power series expansion of the exact result (\ref{er}) in powers of $\theta$ and $\varphi$
retaining just the first order
\begin{equation}
\bm{u}_0 (s) \simeq \left [ \sigma_0 + i \theta \sigma_z +  \varphi \frac{\sin \delta}{\delta} 
\left ( \sigma_+ e^{-i \delta} - \sigma_- e^{i \delta} \right ) \right ] \bm{u}_0 (0) ,
\end{equation}
where $\sigma_0$ is the identity matrix, and $\sigma_\pm = \sigma_x \pm i \sigma_y$. 

We can appreciate that after an helix pith of arc-length $s_0$ with $\Omega s_0 =  2 \pi \sqrt{1 + \Omega^2 r_0^2}$,
the spin $\bm{u}_0 (s_0)$ does not return to its original state $\bm{u}_0 (0)$, but 
\begin{equation}
\label{g1}
\bm{u}_0 (s) = \left ( \sigma_0 + i \theta \Omega \sigma_z \right ) \bm{u}_0 (0) \simeq   e^{ i \theta  \sigma_z } \bm{u}_0 (0) ,
\end{equation}
having taken into account that for $s = s_0$ we have $\delta = \pi$. This is a simple small rotation with axis 
$z$ (see Fig. 5), with angle
\begin{equation}
\label{g2}
\theta \simeq \frac{\pi k r_0^2}{mc^2}.
\end{equation}

\begin{figure}
\includegraphics[width=6cm]{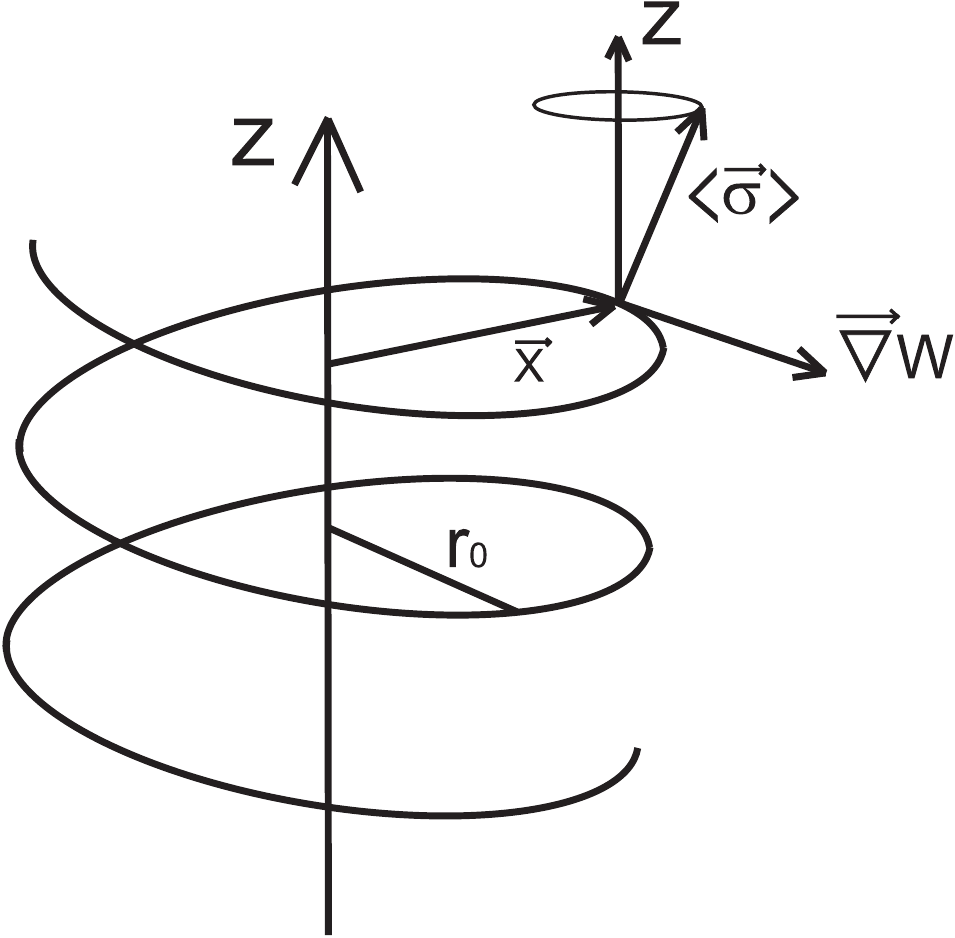}
\caption{Diagram illustrating the evolution of the spin state represented by the vector $\langle \bm{\sigma} 
\rangle$ rotating around the axis  $z$ which is the axis of the helix.}
\end{figure}

This result strongly recalls the geometric phase arising in the optical case of polarization transport along 
an helix \cite{opt}, with also results in a rotation of polarization around the axis $z$ of the Poincar\'e sphere 
(circular polarization being at the poles). However, there is an strong difference since in our case 
the spin transformation after an helix pitch does not depend just on the form of the trajectory, given by
 $\Omega$ and $r_0$. Besides, it depends also on dynamical features  represented by the factor $k$ 
 depending on the particular potential leading to such trajectory. 

We can try to develop further the optical-mechanical analogy by considering that the central potential
is a harmonic oscillator $V= k (x^2 + y^2)/2$. Denoting by $v_z$ the speed along the helix axis we have 
$z= v_z t$ and $\Omega z = \Omega v_z t = \sqrt{k/m} \, t$ so that $k = m \Omega^2 v_z^2$. After Eqs. (\ref{g1})
and (\ref{g2})  the angle rotated by the spin after an helix pitch is
\begin{equation}
\theta \simeq \pi \Omega^2 r_0^2 \frac{v_z^2}{c^2} .
\end{equation}
This can be compared with the angle rotated by light polarization per pitch for the same helix, which is 
$2 \pi /\sqrt{1+ \Omega^2 r_0^2}$. We appreciate that the dependence on $\Omega r_0$ is different,
and that in the case of matter and that in the mechanical case there is a factor $v_z / c$, which is typically  
small.

\section{Conclusions}

We have studied the behaviour of the spin state of a 1/2-spin particle when its spinorial wave function is 
transported along classical trajectories. As it happens in classical electromagnetism, classical trajectories
can emerge within a wave-like picture via two procedures: from the exact propagation of discontinuities and 
from the eikonal approximation. 

It is known that the propagation of discontinuities for the Dirac equation with purely scalar potentials (i.e., 
the analogous situation for the light in isotropic and inhomogeneous media) does not provide enough 
information about the spin of the particle. Then, a first option is introducing an explicitly relativistic implementation 
of a classical scalar potential involving spin-dependent terms: the Dirac oscillator.  This approach has shown a 
definite relation between the spin discontinuity and the tangent to the classical trajectory, being an exact solution.

For an arbitrary scalar potential, we cannot obtain exact results for the spin propagation. Thus we turn 
our attention to the approximation picture provided by WKB methods, establishing as many clear parallels as 
possible with the equivalent and known situation for the electromagnetic field. In this way we have been 
able to find the equation of the spin transport for Dirac particles.  This result reveals three main differences 
regarding the Maxwell asymptotic approach: (i) the electromagnetic field satisfies orthogonality relations (\ref{Me1}) 
that are absent in the Dirac  case, (ii) there is a mass term difference in the Dirac eikonal (\ref{Wf}) in comparison 
with the Maxwell case (\ref{eik}), (iii) in the Dirac case there is no projection of spin along the tangent to the trajectory 
in Eq. (\ref{dsu0}) at difference with the Maxwell case (\ref{dsu0M}). These differences might explain the lack of 
spin-propagation effects in the non relativistic limit, and the lack of topological features in the matter case. In 
particular, we have applied this theory to two non-trivial cases: circular and helicoidal trajectories, providing 
two simple examples for the general results of the main discussion.  
 
\appendix

\section{Eikonal approach for the electromagnetic field}

We review here the eikonal approach for electromagnetic field to be compared 
with the Dirac case. The Maxwell's field equations for time-harmonic waves of frequency
$\omega$ within isotropic and inhomogeneous media are
\begin{equation}
\label{Me1}
 \nabla \left [ \epsilon (\bm{x} ) \bm{E} (\bm{x},t) \right ] =0, \qquad
\nabla \left [ \mu (\bm{x} ) \bm{H} (\bm{x},t) \right ] =0 ,
\end{equation}
\begin{eqnarray}
\label{Me2}
& -\frac{i}{k_0} \nabla \times  \bm{E} (\bm{x},t) =  \mu (\bm{x} ) c \bm{H} (\bm{x},t) , & \nonumber \\
 & & \nonumber \\
& \frac{i}{k_0} \nabla \times  \bm{H} (\bm{x},t) =  \epsilon  (\bm{x} ) c \bm{E} (\bm{x},t)  ,&
\end{eqnarray}
where $k_0 = \omega/c$. They can be combined to get the vectorial wave equation
\begin{equation}
\label{we}
 \nabla^2 \bm{E} + n^2 k_0^2 \bm{E} +  \nabla \left ( \frac{1}{\epsilon} \nabla \epsilon \cdot 
 \bm{E} \right ) + \frac{1}{\mu} \nabla  \mu \times \left ( \nabla \times \bm{E} \right )=0 ,
\end{equation}
where $n(\bm{x} ) = c \sqrt{\epsilon (\bm{x} ) \mu (\bm{x} )}$ is the refraction index.

To apply the eikonal approximation in the limit $1/k_0 \rightarrow 0$ we consider asymptotic  
solutions of the form \cite{KO}
\begin{equation}
\label{expanE}
\bm{E} = \left ( \bm{E}_0 +\frac{1}{k_0}  \bm{E}_1 + \ldots \right ) e^{i k_0 S^\prime} ,
\end{equation}
where $S^\prime (\bm{x} ,t) = L (\bm{x} )-c t$ is the phase. Introducing Eq. (\ref{expanE}) 
into Eq. (\ref{we})  we get for the lowest $1/k_0^0$ order
\begin{equation}
\label{eik}
\left [ \nabla L (\bm{x} ) \right ]^2 =n^2 (\bm{x}) ,
\end{equation}
which is the well known eikonal equation. 

For the $1/k_0^1$  order we have
\begin{equation}
 \label{et}
\frac{d}{ds} \bm{E}_0 = - \frac{\nabla^2 L}{2 | \nabla L | } \bm{E}_0
- \frac{\left ( \bm{E}_0 \cdot \nabla \epsilon \right) \nabla L}{2 \epsilon | \nabla L | }
- \frac{\nabla \mu \times \left ( \nabla L \times \bm{E}_0 \right )}{2 \mu | \nabla L |}  ,
\end{equation}
where 
\begin{equation}
\frac{d}{ds} = \frac{ \bm{\nabla} L \cdot \bm{\nabla}}{\left | \bm{\nabla} L \right | } ,
\end{equation}
is the derivation with respect the arc length  $s$. The last term in Eq. (\ref{et}) can be 
expressed also as 
\begin{equation}
 \label{}
 \frac{\nabla \mu \times \left ( \nabla L \times \bm{E}_0 \right )}{2 \mu | \nabla L |}  = 
 \frac{\left ( \bm{E}_0 \cdot \nabla \mu \right) \nabla L}{2 \mu | \nabla L | } -
  \frac{\left ( \nabla L \cdot \nabla \mu \right) \bm{E}_0 }{2 \mu | \nabla L | } .
\end{equation}
This is the transport equation that expresses the propagation of amplitude and  polarization 
of the photon along the trajectories determined by the eikonal equation (\ref{eik}). 

On the other hand, if we insert Eq. (\ref{expanE}) directly in Eq. (\ref{Me1}) we get again  
Eq. (\ref{eik}) plus the explicit relations of orthogonality 
\begin{equation}
\bm{E}_0 \cdot \nabla L = 0, \qquad \bm{H}_0 \cdot \nabla L = 0 .
\end{equation}
The transport equation (\ref{et}) can be further split into  propagation equations for the 
local amplitude $ | \bm{E}_0  |$ and the local polarization state $\bm{u}_0$, both 
depending on $\bm{x}$,  after decomposing $\bm{E}_0$ in the form 
\begin{equation}
\bm{E}_0 = | \bm{E}_0  | \bm{u}_0 , \qquad | \bm{u}_0  | = 1 .
\end{equation}
This  leads to the following equation of propagation for the local amplitude  $ | \bm{E}_0  |$ 
\begin{equation}
\frac{d}{ds} | \bm{E}_0 | = - \left ( \frac{\nabla^2 L}{2 | \nabla L | } - 
 \frac{ \nabla L \cdot \nabla \mu  }{2 \mu | \nabla L | } \right ) | \bm{E}_0 | ,
\end{equation}
and to this one for the local polarization state $\bm{u}_0$
\begin{equation}
\label{dsu0M}
\frac{d}{ds} \bm{u}_0 = - \frac{\left ( \bm{u}_0 \cdot \nabla \epsilon \right) \nabla L}{2 \epsilon | \nabla L | }
-  \frac{\left ( \bm{u}_0 \cdot \nabla \mu \right) \nabla L}{2 \mu | \nabla L | }   .
\end{equation}
  
\section*{Acknowledgments}

A. L. acknowledges support from projects FIS2012-35583 of 
the Spanish Ministerio de Econom\'{\i}a y Competitividad and QUITEMAD S2009-ESP-1594 of the 
Consejer\'{\i}a de Educaci\'{o}n de la Comunidad de Madrid.

\end{document}